\documentclass[10pt,aps,prb,twocolumn,floatfix,floats,showpacs,superscriptaddress,raggedbottom]{revtex4-1}
\usepackage{graphicx,latexsym}
\usepackage{dcolumn}
\usepackage{amsmath}
\usepackage{amssymb,bm}
\usepackage{color}
\usepackage[normalem]{ulem}

\def\mb#1{\mbox{\boldmath$#1$}}

\def\eq#1{Eq.\ (\ref{#1})}
\def\fig#1{Fig.\ \ref{#1}}

\usepackage{hyperref}
\hypersetup{
    pdfnewwindow=true,       
    colorlinks=true,         
    linkcolor=blue,          
    citecolor=blue,          
    filecolor=magenta,       
    urlcolor=black           
}

\begin{document}

\title{Photon-induced tunability of the thermospin current in a Rashba ring}

\author{Nzar Rauf Abdullah}
\email{nzar.r.abdullah@gmail.com}
\affiliation{Physics Department, College of Science, 
             University of Sulaimani, Kurdistan Region, Iraq}
\affiliation{Komar Research Center, Komar University of Science and Technology, Sulaimani City, Iraq} 
\affiliation{Science Institute, University of Iceland, 
             Dunhaga 3, IS-107 Reykjavik, Iceland}
             
\author{Thorsten Arnold}
\affiliation{Center for Advancing Electronics Dresden,
Dresden University of Technology,
01062 Dresden, Germany}             
             
\author{Chi-Shung Tang}
\affiliation{Department of Mechanical Engineering,
  National United University, 1, Lienda, Miaoli 36003, Taiwan} 
  
\author{Andrei Manolescu}
\affiliation{Reykjavik University, School of Science and Engineering,
              Menntavegur 1, IS-101 Reykjavik, Iceland}    
             
\author{Vidar Gudmundsson}    
\email{vidar@hi.is}
\affiliation{Science Institute, University of Iceland, 
             Dunhaga 3, IS-107 Reykjavik, Iceland}


\begin{abstract}
The goal of this work is to show how the thermospin
polarization current in a quantum ring changes in the presence of 
Rashba spin-orbit coupling and a quantized single photon mode
of a cavity the ring is placed in. Employing the reduced
density operator and a general master equation formalism, we find that
both the Rashba interaction and the photon field can significantly
modulate the spin polarization and the thermospin polarization current.
Tuning the Rashba coupling constant, degenerate energy levels are
formed corresponding to the Aharonov-Casher destructive phase interference in the
quantum ring system. Our analysis indicates that the maximum spin
polarization can be observed at the points of degenerate energy levels
due to spin accumulation in the system without the photon field.
The thermospin current is thus suppressed. In the presence of the
cavity, the photon field leads to an additional kinetic momentum of the
electron. As a result the spin polarization can be enhanced by the photon field.


%
\end{abstract}




\maketitle

\section{Introduction}

Two-dimensional electron gas (2DEG) systems with spin-orbit interactions 
present fascinating properties of quantum transport due to 
the combination of both spin and orbital degrees of freedom.  
This behavior realizes a possibility to control the thermal spin in spintronics~\cite{PhysRevB.83.075310,Lai2015}
and caloritronics~\cite{Bart_Nature_2012}.
In general, the research field of thermoelectricity of nanoscale systems is divided into two directions. 
The first is spin caloritronics~\cite{Bart_Nature_2012,Maekawa_2008}, that studies thermoelectricity in
magnetic structures and spintronic devices. The second is about the 
influences of electromagnetic field on heat transport, which investigates thermoelectricity
under photon radiation at very low temperatures~\cite{PhysRevLett.100.155902,PhysRevB.83.125113}.

In spintronic devices, the temperature gradient-induced spin polarization
recently became an active research field~\cite{PhysRevB.89.155432}, with emerging 
spin Nerst effects~\cite{PhysRevB.87.075301}.
The Rashba spin-orbit interaction is widely present in the physics of 
spin thermoelectric devices because of fascinating properties of 
nontrivial topology of Fermi surfaces~\cite{PhysRevLett.98.167002,Xiao2016}
that has been detected in several interesting investigations, for example, 
the controlled thermoelectric figure of merit 
and the spin relaxation behavior~\cite{Lidong_APL_2014,PhysRevB.72.075307}.
It was expected that the above mentioned behavior of the Rashba interaction 
would not survive in narrow gap semiconductor because of the small 
Rashba spin splitting energy~\cite{PhysRevLett.78.1335}. 
But it has been shown that the Rashba interaction has a major role
to enable control of spin-dependent thermoelectric properties in semiconductor devices~\cite{Lidong_APL_2014,Abdullah2017}.

The influences of a photon field on thermal-spin properties of nanoscale systems have 
been studied using  nonequilibrium Green’s function theory. It was shown 
that the photon field enhances the spin-dependent Seebeck coefficients and the figures of merit~\cite{PhysRevB.87.085427}.
Additionally, the interplay between the photon field and the thermally induced
electron populations results in a switching of the thermal-current sign as time evolves and its stationary value can be
maximized by tuning the laser intensity~\cite{0953-8984-25-13-135301}.
We have recently shown that the thermoelectric current can be controlled by a quantized cavity photon field 
where the photon number and the photon polarization play a major role to magnify thermoelectric 
currents in a quantum wire~\cite{Nzar_ACS2016}.
Furthermore, spin-dependent heat and thermoelectric currents influenced by
the Rashba interaction were presented~\cite{Abdullah2017}. 
The heat current is suppressed in the presence of the photon cavity due 
to the contribution of the two-electron and photon replica states to the transport 
while the thermoelectric current is not sensitive to changes in the parameters of the photon field.

In the present work, we show that the spin-polarization and thermospin current in a quantum ring coupled 
to a photon cavity can be controlled by a Rashba spin-orbit coupling and the electron-photon coupling strength.

The paper is organized as follows: In Sec.~\ref{Sec:II}, we present
the model describing a quantum ring coupled to a photon cavity. 
Section \ref{Sec:III} shows the numerical results and
discussion. Concluding remarks are addressed in Sec.~\ref{Sec:IV}.

%

\section{Model and Theory}\label{Sec:II}

\begin{figure}[htb]
\centering
    \includegraphics[width=0.45\textwidth,angle=0]{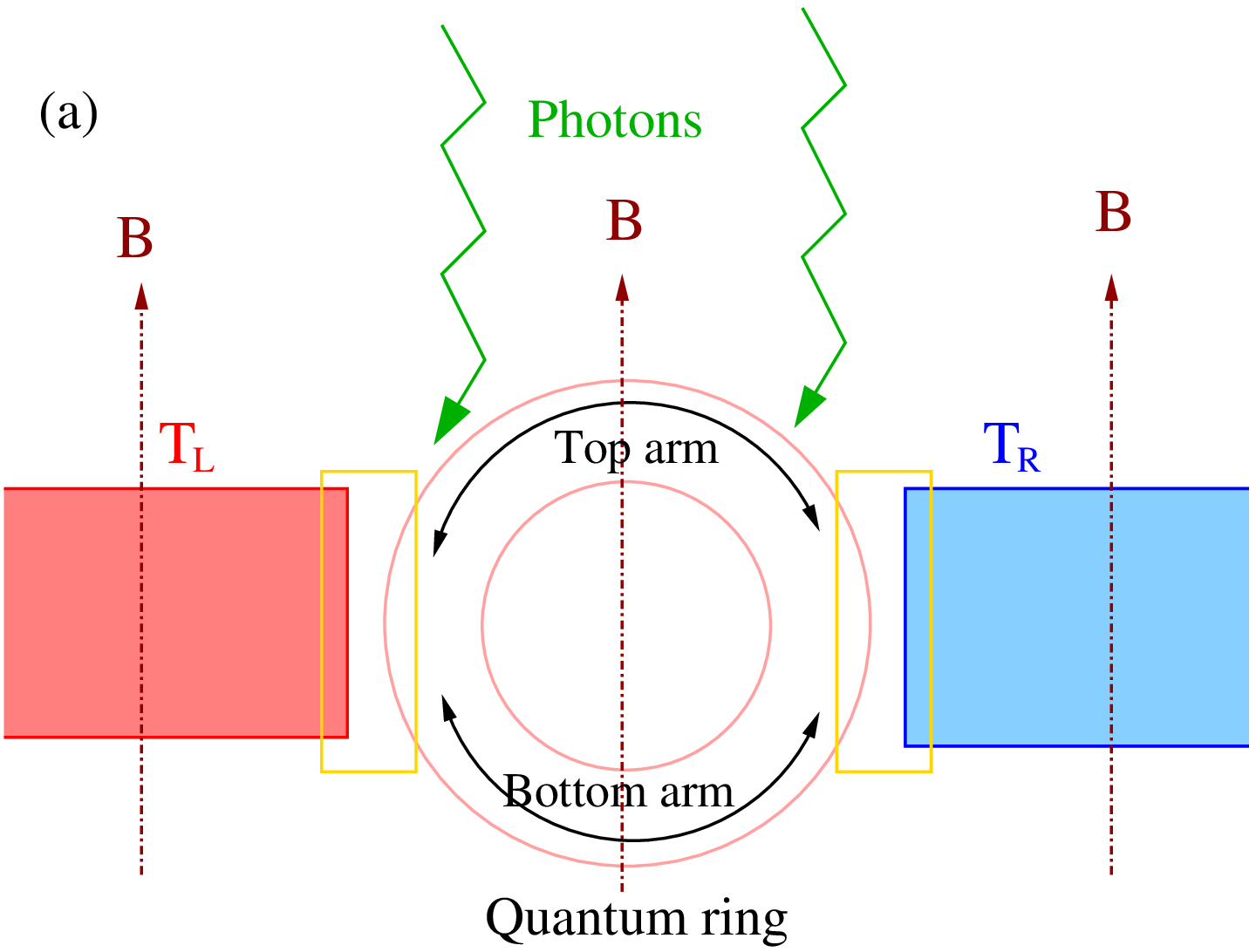}
    \includegraphics[width=0.45\textwidth,angle=0]{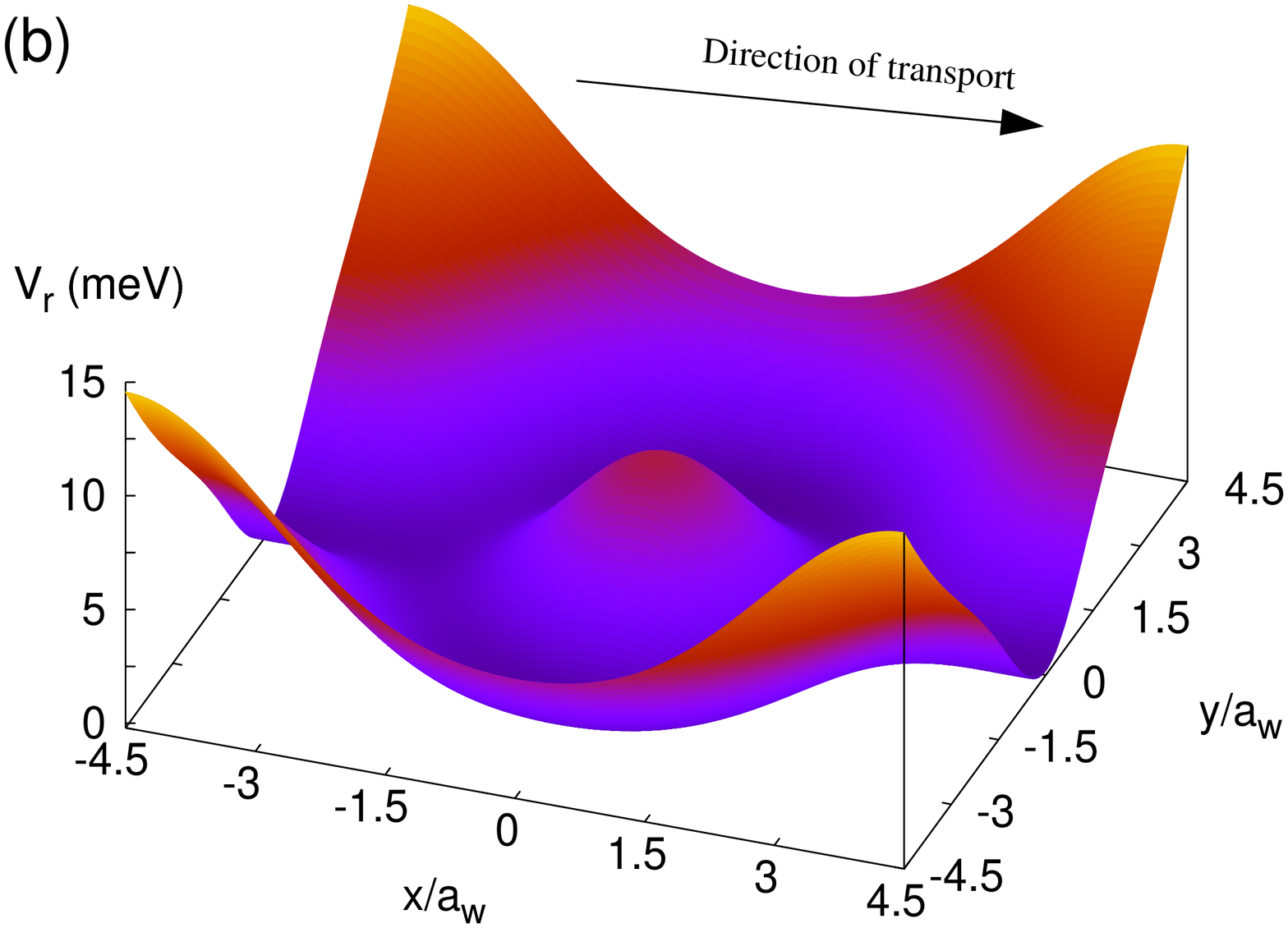}
 \caption{(Color online) (a) Schematic diagram showing the QR (pink color) connected 
 to the leads, where the temperature of the left lead ($T_L$) (red color) is higher than the temperature of the 
 right lead ($T_R$) (blue color). The golden color indicates the contact regions between the QR and the leads.
 The brown arrows indicate the external magnetic fields and the 
 green zigzag display the photon field in the cavity. (b) The potential $V_r(\mathbf{r})$ defining the central ring
       system that will be coupled diametrically to the semi-infinite left and right leads
       in the $x$-direction.}
\label{fig01}
\end{figure}

In this section, we introduce our model, 
the Hamiltonian of the system and the formalism to calculate the
thermospin current. The system consists of a quantum ring (QR) connected 
to two leads with different temperatures 
and the total system is exposed to a small external perpendicular magnetic field.
The QR system is coupled to a photon cavity with a single quantized linearly polarized photon mode. 
The QR is considered to be built on a two-dimensional electron gas of a GaAs/AlGaAs
material in the $xy$-plane.
The QR embedded in the central system with length $L_x = 300$~nm
is schematically shown in \fig{fig01}(a) and the potential $V_r(\mathbf{r})$ defining 
the central ring system that will be coupled diametrically to the semi-infinite 
left and right leads in the $x$-direction.

The QR system coupled to the photon cavity can be described 
by a Hamiltonian of
the form 

\begin{eqnarray}
\hat{H}_{S}&=& \int d^2 r\; \hat{\mathbf{\Psi}}^{\dagger}(\mathbf{r})\Big[\frac{1}{2m^{*}} 
\Big( \frac{\hbar}{i}\nabla +\frac{e}{c} \big(\hat{\mathbf{A}}_B(\mathbf{r}) +
\hat{\mathbf{A}}_{\gamma}(\mathbf{r})\big)   \Big)^2 \nonumber \\
&& +  V_r(\mathbf{r}) + H_{Z} 
 + \hat{H}_{R}(\mathbf{r}) \Big] \hat{\mathbf{\Psi}}(\mathbf{r}) 
 +\hat{H}_{ee}+\hbar\omega_{\gamma} \hat{a}^{\dagger}\hat{a}. \label{H^S}
\end{eqnarray}

Herein, the magnetic vector potential is $\hat{\mathbf{A}}_B(\mathbf{r}) = -By\hat{x}$ with 
$\mathbf{B} = B \hat{\mb{z}}$, and  the photon vector potential is 
$\hat{\mathbf{A}}_{\gamma}(\mathbf{r}) = A(\mathbf{e}\hat{a}+\mathbf{e}^{*}\hat{a}^{\dagger})$, 
which is defined in terms of photon creation ($\hat{a}^{\dagger}$) and annihilation ($\hat{a}$) operators
with $\mathbf{e}= \mathbf{e}_x$ for the longitudinal photon polarization ($x$-polarization).
The strength of the 
photon vector potential $A$ determines the electron-photon coupling constant $g_{\gamma} = eA\Omega_w a_w/c$, where 
$\Omega_w = (\omega^2_c + \Omega^2_0)^{\frac{1}{2}}$ is the effective characteristic frequency 
with $\Omega_0$ the frequency of the confined electron in the $y$-direction, $\omega_c$ is 
the cyclotron frequency and $a_w$ the effective magnetic length.

In the second line of \eq{H^S}, $V_r(\mathbf{r})$ is the confinement potential of the ring~\cite{Arnold2014}, 
$H_{Z} = \frac{1}{2} (\mu_B g_S B \sigma_z)$
is the Zeeman spin effect with $\mu_B$ the Bohr magnetron and $g_S$ is
the electron spin g-factor and $\hat{H}_{R}(\mathbf{r})$ is the Rashba-spin orbit coupling that defines 
the interaction between the orbital motion and the spin of an electron
\begin{equation}
 \hat{H}_{R}(\mathbf{r})=\frac{\alpha}{\hbar}\left( \sigma_{x} \hat{p}_y(\mathbf{r}) -\sigma_{y} \hat{p}_x(\mathbf{r}) \right).
 \label{H_R} 
\end{equation}
Herein, $\alpha$ indicates Rashba coupling constant and 
$\sigma_{x}$ and $\sigma_{y}$ are the Pauli matrices.
The exact diagonalization techniques is utilized to calculate the 
electron-electron Coulomb interaction ($\hat{H}_{ee}$) 
in the QR system~\cite{PhysRevB.82.195325,Nzar_IEEE_2016,Nzar.25.465302,ABDULLAH2016280}.
The free photon in the cavity is described by the Hamiltonian $\hbar \omega_{\gamma}\hat{a}^{\dagger}\hat{a}$, where
$\hbar \omega_{\gamma}$ is the photon energy.

In addition, the electron spinor field operators are 
\begin{equation}
      \hat{\mathbf{\Psi}}(\mathbf{r})= \left( \begin{array}{c} \hat{\Psi}(\uparrow,\mathbf{r}) \\ 
      \hat{\Psi}(\downarrow,\mathbf{r}) \end{array} \right), \quad 
      \hat{\mathbf{\Psi}}^{\dagger}(\mathbf{r})=
      \left(\begin{array}{cc} \hat{\Psi}^{\dagger}(\uparrow,\mathbf{r}), & \hat{\Psi}^{\dagger}(\downarrow,\mathbf{r}) \end{array}\right), 
      \label{conj_FOS}
\end{equation}
where $\hat{\Psi}(x)=\sum_{a}\psi_{a}^{S}(x)\hat{C}_{a}$ is the field operator with 
$x\equiv (\mathbf{r},\sigma)$, $\sigma \in \{ \uparrow,\downarrow \}$ and the annihilation operator, $\hat{C}_{a}$,
for the single-electron state (SES) $\psi_a^S(x)$ in the QR system.

To describe the transient electron transport through the QR system, we use 
a time-convolutionless generalized master equation (TCL-GME)~\cite{Arnold13:035314,Arnold2014,Thorsten2014}.
The TCL-GME is local in time and satisfies the positivity for the many-body state occupation described by the reduced density operator (RDO). 
The total density matrix $\hat{\rho}_T$, before the QR system is coupled to the leads, is the product of the 
density matrix of the QR system and the leads. From the
total density matrix  
the reduced matrix can be defined after the coupling of the QR system to the leads.
The RDO of the system QR system $\hat{\rho}_S$ is thus expressed as 
\begin{equation} 
 \hat{\rho}_S(t) = {\rm Tr}_l \big[\hat{\rho}_T(t) \big] ,
 \label{RDO}
\end{equation}
where $l \in \{L,R\}$ refers to the two electron reservoirs, the left (L) and the right (R) leads, respectively.

The coupling of the QR system and the leads is defined by a
many-body coupling tensor derived from the geometry of the single-electron states (SES) in the
contact area of the leads and system~\cite{Vidar61.305,PhysRevB.81.155442}. 
Details about all parameters for the coupling are found in~\cite{Arnold13:035314}. 
The time needed to reach the steady state
depends on several parameters such as the chemical potentials of both leads, the bias
window, its relation to the energy spectrum of the
system, and the nature of the relevant many-body states. 
In our calculations we integrate the GME to $t = 220$ ps, a
point in time late in the transient regime when the system
is approaching the steady state.

In our present work, we calculate some physical quantities that can explain the spin-dependent 
thermal properties of the QR system.
We first introduce the spin polarization of the QR system
%
%
in $i=x,y,z$ direction $S_i$ and its spin polarization operator
\begin{equation}
 \hat{S}_i = \int d^2r \hat{n}^i(\mathbf{r}) ,
\end{equation}
where $\hat{n}^i(\mathbf{r})$ is the spin polarization density operator for the spin polarization 
$S_i$~\cite{ARNOLD2014170}. 
To further investigate  thermal properties of the QR system, 
we introduce the top local thermospin current ($I^{{\rm th},i}_{\rm t}$) through the upper arm ($y>0$) of the QR system 
\begin{equation}
 I^{{\rm th},i}_{\rm t}(t) = \int_{0}^{\infty} dy \, j_x^{{\rm th},i}(x=0,y,t) 
\end{equation}
and the bottom local thermospin polarization current ($I^{{\rm th},i}_{\rm b}$) through the lower arm ($y<0$) of the QR system
\begin{equation}
 I^{{\rm th},i}_{\rm b}(t) = \int_{-\infty}^{0} dy \, j_x^{{\rm th},i}(x=0,y,t) ,
\end{equation}
where the spin polarization current density 
is 
\begin{equation}
 \mathbf{j}^{th,i}(\mathbf{r},t) = {{j^{th,i}_x(\mathbf{r},t)}\choose{j^{th,i}_y(\mathbf{r},t)}} 
 = {\rm Tr} \Big[\hat{\rho}_s(t) \hat{\mathbf{j}}^{th,i}(\mathbf{r}) \Big]
\end{equation}
that is calculated by the expectation value of the spin polarization 
current density operator~\cite{ARNOLD2014170}.

From both the top and the bottom thermospin current polarizations, one can define
the total local (TL) thermospin polarization current
\begin{equation}
 I_{tl}^{{\rm th},i}(t) =  I^{{\rm th},i}_{\rm t}(t) + I^{{\rm th},i}_{\rm b}(t) .
\end{equation}

The TL thermospin polarization current is 
related to non-vanishing spin-polarization sources, 
and the circular local (CL) thermospin polarization current 
is 
\begin{equation}
 I_{cl}^{{\rm th},i}(t) = \frac{1}{2} \Big[ I^{{\rm th},i}_{\rm b}(t) - I^{{\rm th},i}_{\rm t}(t)\Big].
\end{equation}

In the next section, we shall show the main results of the spin polarization and thermospin polarization current 
in the QR system and the influence of the photon field on the QR system \cite{Vidar61.305,PhysRevB.81.155442}. 

\section{Results}\label{Sec:III}

The model used in our calculations consists of a semiconductor QR system coupled to a photon cavity 
with a single photon mode. The QR system is hard wall confined in the $x$-direction and 
parabolically confined in the $y$-direction with characteristic energy $\hbar \Omega_0 = 1.0$~meV.
The potential of the QR is defied by 

\begin{align}
 V_r(\mathbf{r})&= \sum_{i=1}^{6}V_{i}\exp
 \left[
 -\left(\gamma_{xi}(x-x_{0i})\right)^2 - \left(\gamma_{yi}\, y\right)^2
 \right] \nonumber \\
 & +\frac{1}{2}m^* \Omega_{0}^2y^2, 
 \label{V_r} 
\end{align}
where $V_{i}$, $\gamma_{xi}$, and $\gamma_{yi}$ are constant values shown 
in our previous publications~\cite{Arnold13:035314,Abdullah2017}. 
The characteristic energy of the electron confinement 
of the central system is expressed by the second term of \eq{V_r}.

The QR system is exposed to a perpendicular magnetic field $B = 10^{-5}$~T to lift the spin degeneracy.
The value of the magnetic field is out of the Aharonov-Bohm (AB) regime because 
the area of the ring structure is $A = \pi a^2 \approx 2 \times 10^4$ nm$^2$ leading to 
a magnetic field $B_0 = \phi_0/A \approx 0.2$~T corresponding to one flux quantum $\phi_0 = hc/e$~\cite{Arnold2014,0953-8984-28-37-375301}.
Therefore, the applied magnetic field is much smaller than $B_0$, orders of magnitudes outside the AB regime.
The photon energy is assumed to be $\hbar \omega_{\gamma} = 0.55$~meV.
Now, we assume the left and the right leads have the same chemical potential ($\mu_L = \mu_R = \mu$),
but are at different temperatures, which induce a thermospin current to flow in the QR system. 

\begin{figure}[htb]
    \includegraphics[width=0.22\textwidth,angle=0]{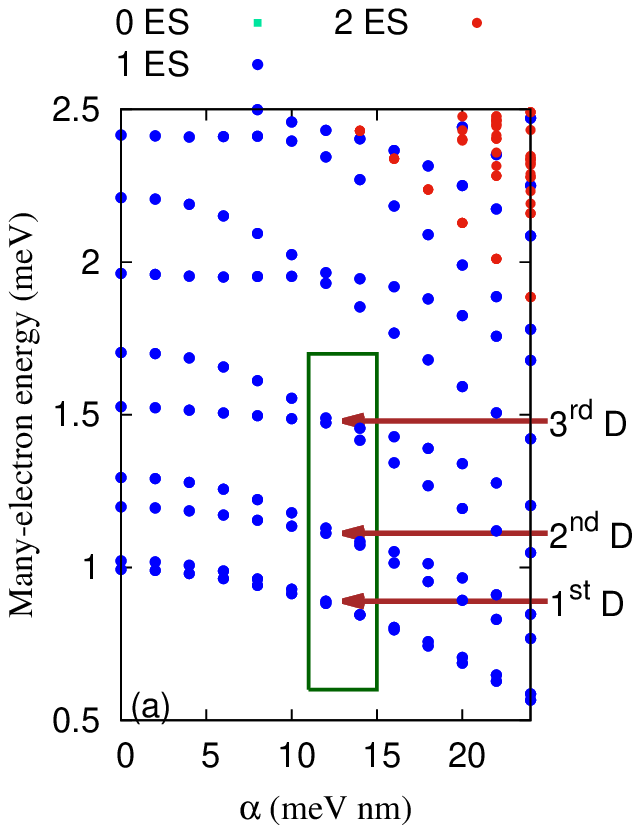}
    \includegraphics[width=0.22\textwidth,angle=0]{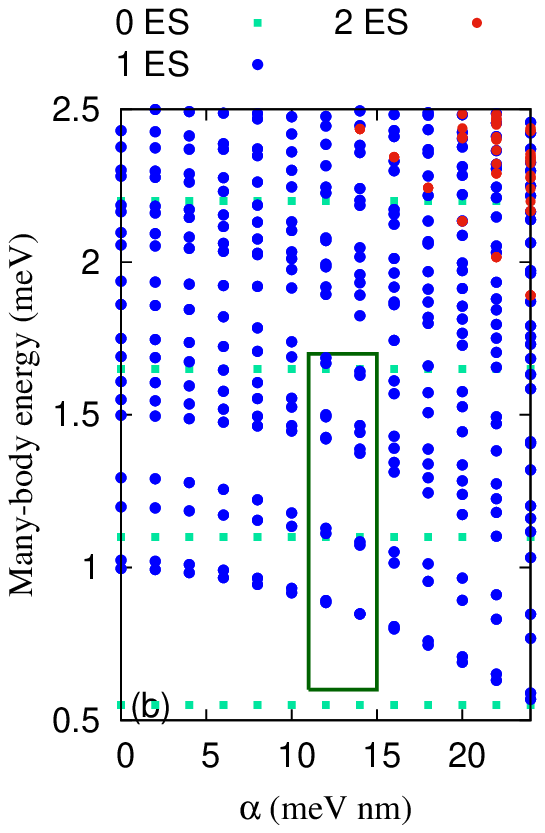}
 \caption{(Color online) Energy spectra of the QR system versus the Rashba coupling constant 
          $\alpha$ without (a), and with (b) a photon cavity. The states 0ES (green rectangles)
          are zero-electron states, 1ES (blue circles) are one-electron states, and 2ES (red circles) are two-electron states.
          The photon energy is $\hbar \omega_{\gamma} = 0.55$~meV, the electron-photon coupling strength
          is $g_{\gamma} = 0.05$~meV, and the photons are linearly polarized in the $x$-direction. 
          The magnetic field is $B = 10^{-5}$~T, and $\hbar \Omega_0 = 1.0~{\rm meV}$.}
\label{fig02}
\end{figure}

In \fig{fig02}, we present the energy spectrum of the QR system without (a), and with (b) 
the photon field. The data reveals a crossing of one-electron states around 
$\alpha = [11\text{-}16]$ meV nm (green rectangle region)
corresponding to the Aharonov-Casher (AC) destructive phase interference~\cite{Arnold2014}. 
The crossing of one-electron states forms the so called 
degenerate energy states. For instance, several degenerate points are formed 
at energy $E_{\mu} = 0.889$, $1.129$ and $1.48$~meV when $\alpha = 12$ meV nm. 
For instance, the first degenerate ($1^{\rm st}$ D) point is between the ground state and the first-excited state while 
the second degenerate ($2^{\rm nd}$ D) point is formed between the second- and third-excited states.
The many-body energy spectrum of the QR in the presence of the photon field is shown in 
\fig{fig02}(b), where the photon field is assumed to be $\hbar \omega_{\gamma} = 0.55$~meV and the 
electron-photon coupling is $g_{\gamma} = 0.05$~meV. Photon replica states are created in the energy spectrum 
of the QR system with the photon field. The photon replica states contribute to the 
electronic transport in the system. The spin polarization and thermo-spin polarization are thus modified by 
the photon field.

\subsection{Spin polarization}

\begin{figure}[htb]
    \includegraphics[width=0.5\textwidth,angle=0]{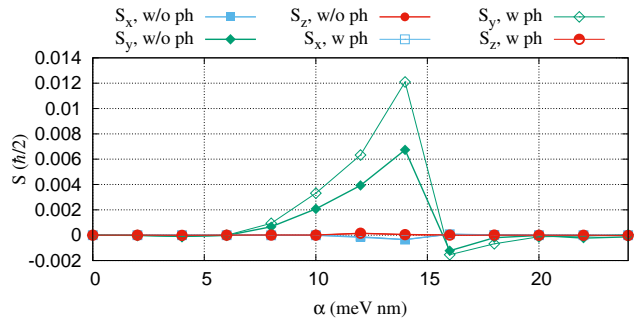}
 \caption{(Color online) Spin polarization $\mathbf{S} = (S_x, S_y S_z)$ of the QR system without 
          (w/o) (solid) and with (w) (hollow) the photon field.  
          The temperature of the left (right) lead is fixed at $T_{\rm L} = 0.41$~K ($T_{\rm R} = 0.01$~K) implying a
         thermal energy $k_B T_{\rm L} = 0.35$~meV ($k_B T_{\rm R} = 0.00086$~meV), respectively.       
          The photon energy is $\hbar \omega_{\gamma} = 0.55$~meV, the electron-photon coupling strength
          is $g_{\gamma} = 0.05$~meV, and the photons are linearly polarized in the $x$-direction. 
          The magnetic field is $B = 10^{-5}$~T, and $\hbar \Omega_0 = 1.0~{\rm meV}$.}
\label{fig03}
\end{figure}

Spin polarization in our system is the degree to which the spin of electrons
is aligned with a given direction. Figure \ref{fig03} indicates the spin polarization 
versus the Rashba coefficient for the system without (w/o) and with (w) the photon field.
We can see that the spin polarization in the $x$- and $z$-directions, $S_x$ and $S_z$, vanishes 
while the spin polarization $S_y$ in the $y$-direction has nonzero value for the Rashba coupling constant around $\alpha=$[$11\text{-}16$]~meV nm
corresponding to the degeneracy of the energy states shown in \fig{fig02}. 
The non-vanishing spin polarization also corresponds to the location of the destructive 
AC interference around $\alpha=$[$11\text{-}16$]~meV. 

In our system, the main transport and the canonical momentum are along the $x$-direction.
Therefore, in the absence of the photon field, the Rashba effective magnetic field should be parallel to the $y$-direction and induce 
a spin polarization in the $y$-direction as is shown in \fig{fig03}. 
But, in the presence of the photon cavity with $x$-polarization, kinetic momentum 
of the electrons in the $x$-direction is added. Consequently, the $S_y$ spin polarization is increased with the 
$x$-polarized photon field (see \fig{fig03}). The large spin polarization around $\alpha= 12$~meV nm (destructive AC phase) 
is due to spin accumulation in the system.

From now on, we only focus on the $y$-component of the spin polarization and thermospin polarization current at
the Rashba coupling constant around $\alpha = [11\text{-}16]$ 
because the spin polarization has nonzero value and high spin polarization is recorded in that range. 

\begin{figure}[htb]
    \includegraphics[width=0.5\textwidth,angle=0]{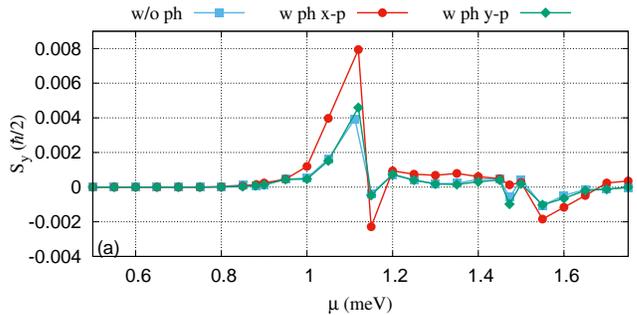}
 \caption{(Color online) $S_y$ spin polarization  of the QR system without 
          (w/o ph) (blue rectangles) and with (w ph) photon field for 
          the $x$-polarization (red circles) and $y$-polarization (green diamonds).
          The temperature of the left (right) lead is fixed at $T_{\rm L} = 0.41$~K ($T_{\rm R} = 0.01$~K) implying a
         thermal energy $k_B T_{\rm L} = 0.35$~meV ($k_B T_{\rm R} = 0.00086$~meV), respectively.       
          The photon energy is $\hbar \omega_{\gamma} = 0.55$~meV, the electron-photon coupling strength
          is $g_{\gamma} = 0.05$~meV. 
          The magnetic field is $B = 10^{-5}$~T, $\alpha = 12$~meV nm, and $\hbar \Omega_0 = 1.0~{\rm meV}$.}
\label{fig04}
\end{figure}

Figure \ref{fig04} shows the $S_y$ spin polarization versus the chemical potential of the leads at 
$\alpha = 12$~meV nm for the QR system without (w/o) (blue rectangles) and with (w) photons in the $x$- (red circles) 
and $y$-polarization (green diamonds). 
In the absence of the photon field, it is clearly seen that the $S_y$ spin polarization 
is maximum at the degenerate energy state around $\mu = 1.12$~meV corresponding to states contributing to the 
AC destructive phase interference. The $S_y$ spin polarization is enhanced in the case of $x$-polarized photon field 
while in the $y$-polarization, the $S_y$ spin polarization remains almost unchanged. 
Therefore, we will only consider the $x$-polarized photon field in our calculations.

\begin{figure}[htb]
    \includegraphics[width=0.5\textwidth,angle=0,bb=50 85 410 230]{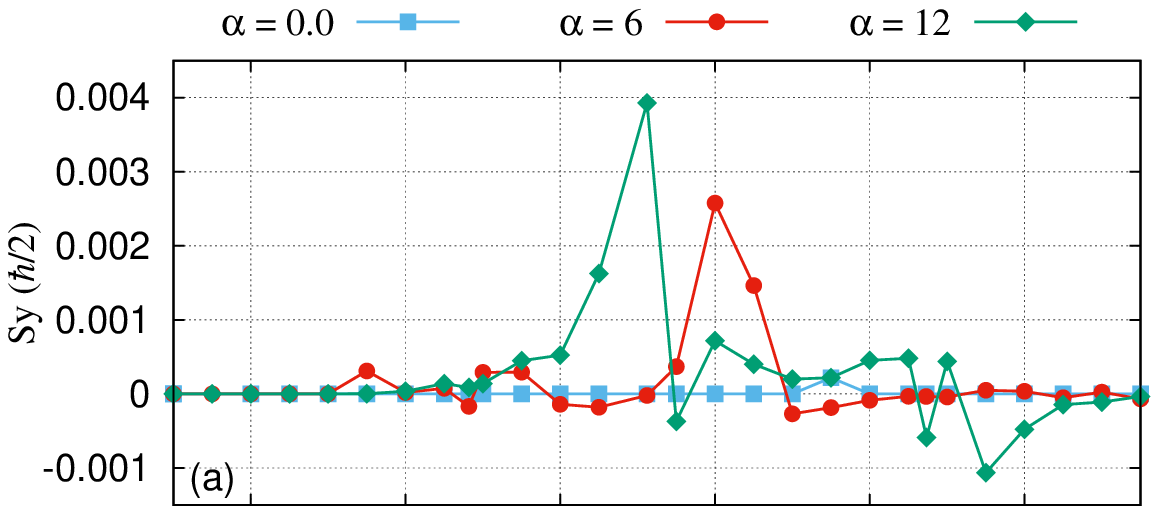}\\
    \includegraphics[width=0.5\textwidth,angle=0,bb=50 85 410 223]{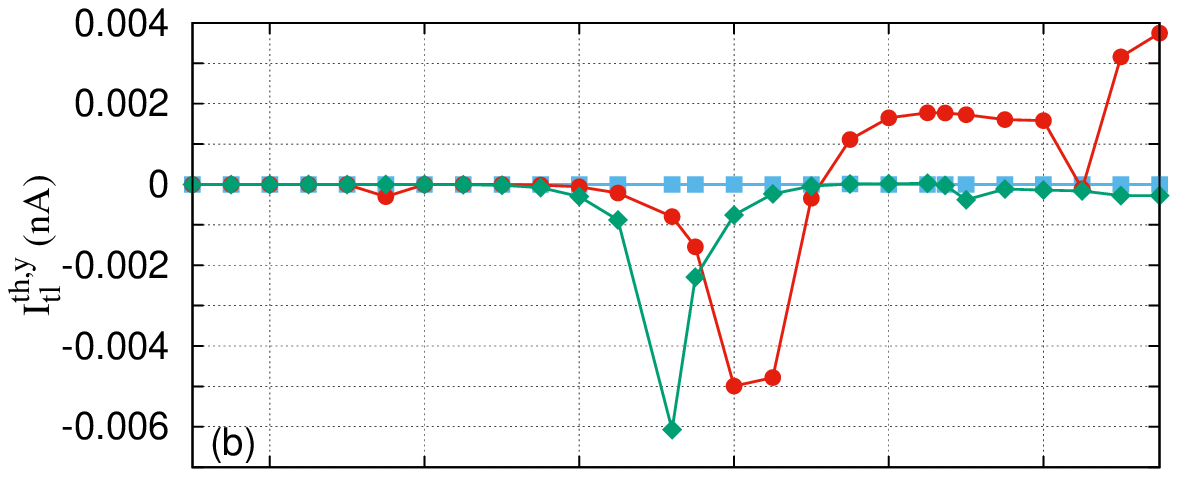}\\
    \includegraphics[width=0.5\textwidth,angle=0,bb=67 60 410 215]{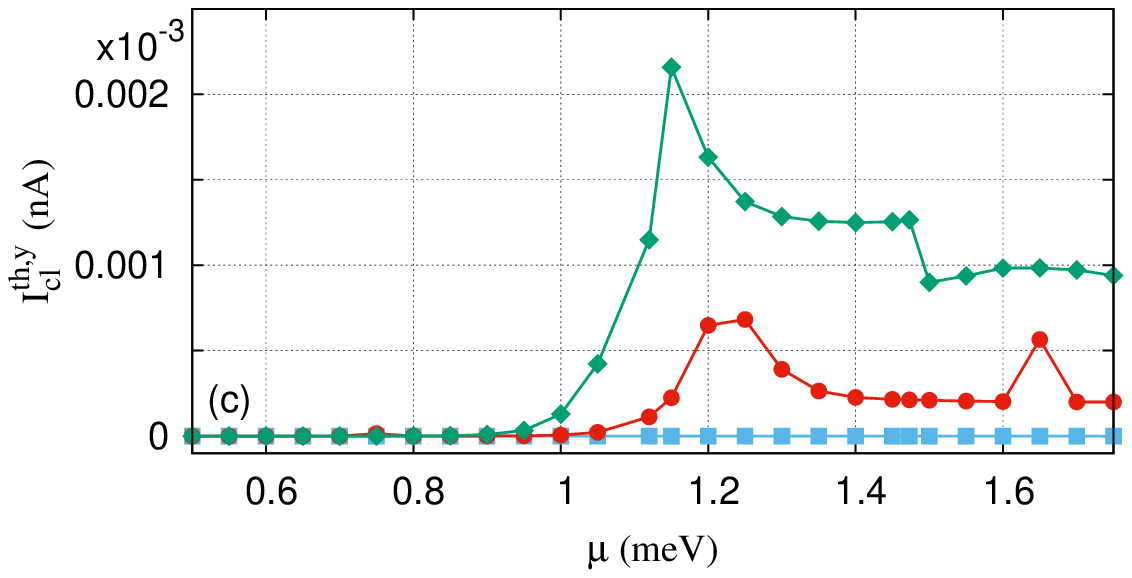}
 \caption{(Color online) (a) $S_y$ spin polarization, (b) TL thermospin polarization current,
 and (c) CL thermospin polarization current versus the chemical potential of 
 the leads $\mu$ for three different values of the 
 Rashba coupling constant $\alpha = 0$~meV nm (blue rectangles),  $\alpha = 6$~meV nm (red circles), 
 and $\alpha = 12$~meV nm (green diamonds) in the absence of the photon field. 
 The temperature of the left (right) lead is fixed at $T_{\rm L} = 0.41$~K ($T_{\rm R} = 0.01$~K) implying 
 thermal energy $k_B T_{\rm L} = 0.35$~meV ($k_B T_{\rm R} = 0.00086$~meV), respectively.             
 The magnetic field is $B = 10^{-5}$~T, and $\hbar \Omega_0 = 1.0~{\rm meV}$.}
\label{fig05}
\end{figure}

Figure \ref{fig05} shows (a) the $S_y$ spin polarization, (b) TL thermospin polarization current ($I^{{\rm Th},y}_{\rm tl}$)
and (c) CL thermospin polarization current ($I^{{\rm Th},y}_{\rm cl}$)
as a function of the chemical potential for different values of the Rashba coupling in the absence of the photon field. 
The $S_y$ spin polarization is zero for $\alpha = 0$~meV nm due to the vanishing spin-orbit interaction and thus 
TL and CL thermospin polarization currents are zero (blue rectangles) for the selected range of the chemical potential of the leads. 
If the strength of the Rashba coupling is tuned to $\alpha = 6$~meV nm, the $S_y$ spin polarization has a peak 
at $\mu = 1.2$~meV. In this case the TL thermospin polarization current 
has a pronounced minimum and the CL thermospin polarization current is slightly enhanced at $\mu = 1.2$~meV (red circles)~\cite{Arnold2014}. 
Increasing the Rashba coupling to $\alpha = 12$~meV nm, which is the critical value of the Rashba coupling corresponding to the 
degenerate energy states shown in \fig{fig02}(a) and describing the position of the destructive AC interference,
the $S_y$ spin polarization is large at $\mu = 1.12$~meV due to spin accumulation at the degeneracy energy of
the $2^{\rm nd}$ D state. Consequently, the TL thermospin polarization current has a minimum value and the CL thermospin polarization current reached its maximum value. The positive value of the CL thermospin current indicates that the direction of electron motion is 
counter-clockwise in the ring~\cite{Arnold2014}. While the maximum absolute value of the CL thermospin current increases strongly with the strength of the Rashba interaction $\alpha$, the TL thermospin current does not increase in the same way. This behavior is because we approach the destructive AC phase.
Another minimum and maximum points of TL and CL thermospin currents, respectively, are observed around $\mu = 1.6$~meV 
corresponding to the higher degenerate energy states.

Both TL and CL thermospin polarization currents are zero 
below $\mu \backsimeq 0.9$~meV because there is no energy state of the ring below $E_{\mu} = 0.9$~meV at low 
spin-orbit coupling strength ($\alpha < 12$ meV nm).
In addition, the first subband of the leads is approximately located at $E_{l} = 1.0$~meV.

\begin{figure}[htb]
    \includegraphics[width=0.5\textwidth,angle=0,bb=50 85 410 230]{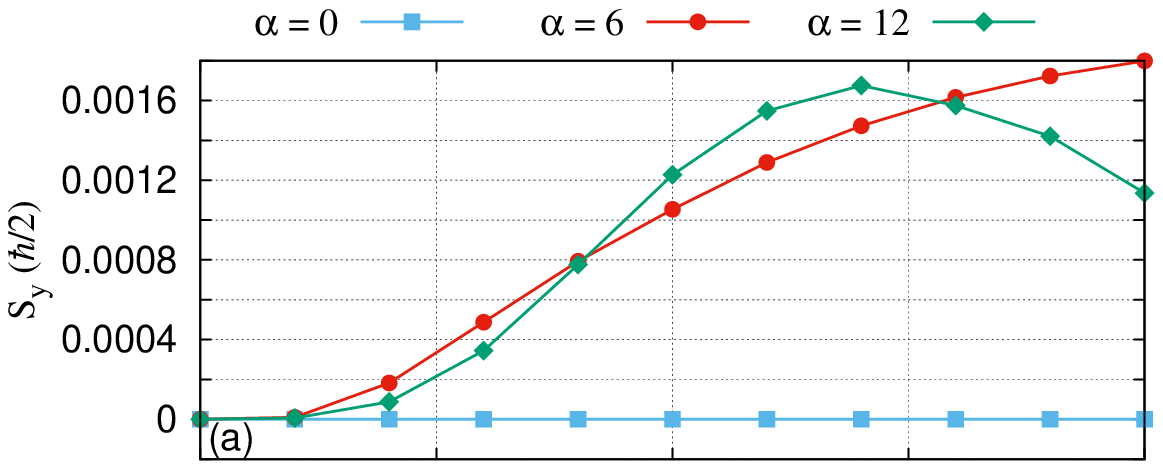}\\
    \includegraphics[width=0.5\textwidth,angle=0,bb=33 75 419 214]{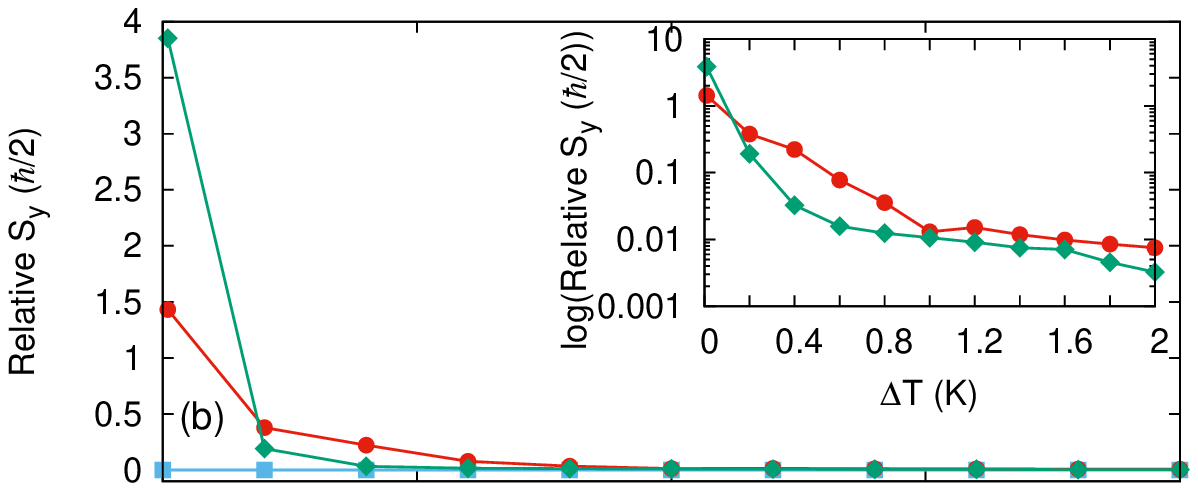}\\
    \includegraphics[width=0.5\textwidth,angle=0,bb=58 75 410 206]{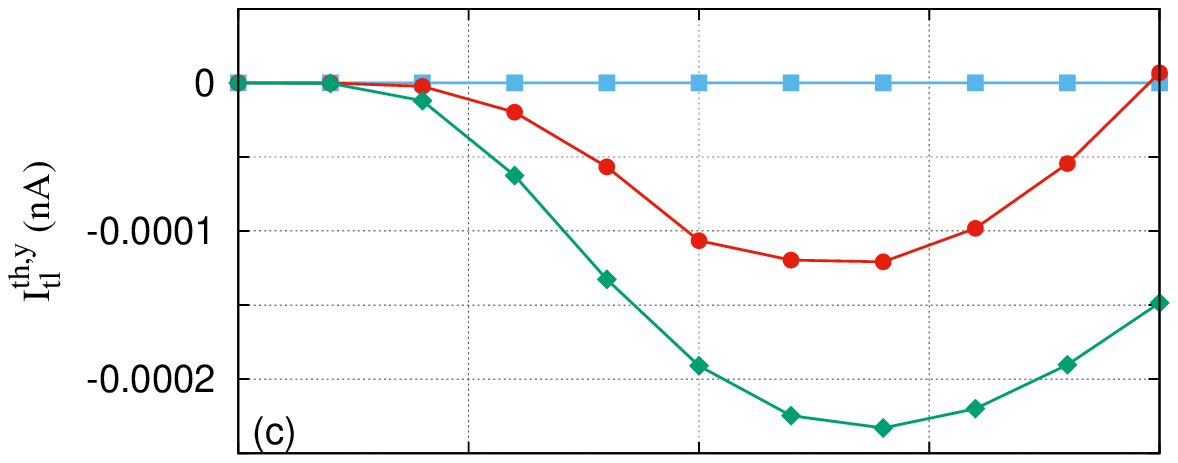}\\
    \includegraphics[width=0.5\textwidth,angle=0,bb=42 60 410 200]{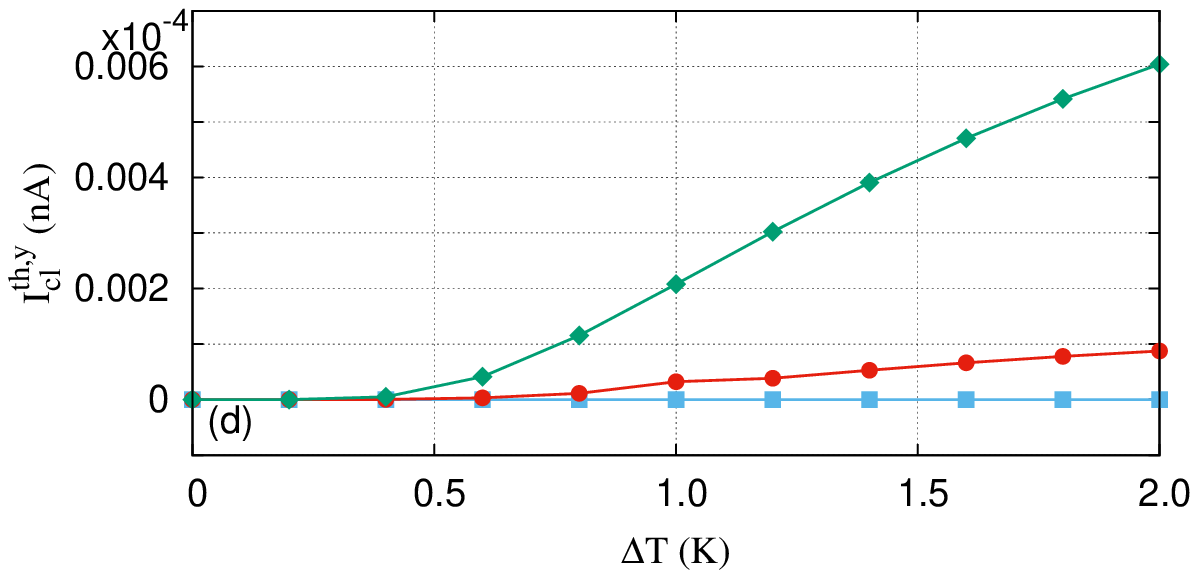}
 \caption{(Color online) (a) $S_y$ spin polarization, (b) $S_y$ spin polarization relative to the number of electrons in the system, (c) TL thermospin polarization current,
 and (d) CL thermospin polarization current versus the temperature gradient $\Delta T$ for three different values of the 
 Rashba coupling constant $\alpha = 0$~meV nm (blue rectangles),  $\alpha = 6$~meV nm (red circles), 
 and $\alpha = 12$~meV nm (green diamonds) in the absence of the photon field. 
 The magnetic field is $B = 10^{-5}$~T, and $\hbar \Omega_0 = 1.0~{\rm meV}$.}
\label{fig06}
\end{figure}

We now tune the temperature of the leads. The temperature of the right lead is fixed at 
$T_R = 0.01$~K and we change the temperature of the left lead. Figure \ref{fig06} shows (a) the 
spin polarization, (b) relative $S_y$ spin polarization, (c) TL thermospin polarization current,
and (d) CL thermospin polarization current versus the temperature gradient of the leads 
for three different values of the Rashba coupling constant at 
$\mu = 1.12$~meV (at the $2^{\rm nd}$ D).
The spin polarization increases by enhancement of the temperature in the presence of the Rashba effect.
That is because the number of electrons in the QR system increases at higher temperature gradient, 
and the spin thus increases. We plot the relative $S_y$ spin polarization, which is the ratio of 
the $S_y$ spin polarization to the number of electrons (sum of occupation numbers) in the QR system [see \fig{fig06}(b)].
It indicates that the relative spin polarization decreases with the temperature gradient.
Thus, the enhancement of the spin polarization with the temperature gradient is only 
due to the enhancement of the number of electrons in the QR system. 
As a result, the TL and CL thermospin polarization is enhanced at high temperature gradient.
The increase could be caused by an improved injection of electrons from the left lead or a stronger 
thermal ``pressure'' of electrons to be injected from the left lead than from the right leading to 
a larger current (without destructive AC phase). Still, the destructive AC phase would hinder 
the electrons from reaching the right lead meaning that more spin-carrying electrons will 
accumulate in the QR system. In addition, with a larger temperature, the energetic matching required for 
a large spin polarization seen from \fig{fig05} might be better fulfilled due to broadening in energy.
The inset of \fig{fig06}(b) shows the logarithmic relative $S_y$ spin polarization.


\subsection{Electron-photon effects}

We consider the case that the QR system is coupled to a photon field with initially empty photon cavity.
In this section, the Rashba coupling constant is fixed at $\alpha = 12$~meV nm corresponding to the 
degenerate energy levels (green rectangle) shown in \fig{fig02}.
The photon energy is $\hbar \omega_{\gamma} = 0.55$~meV which is approximately equal to the 
energy spacing between the $1^{\rm st}$ D and the third degenerate point ($3^{\rm rd}$ D) forming between 
the fourth- and fifth- excited states at $\alpha = 12$~meV. 
Under this condition, the QR system is in resonance with the photon field.

\begin{figure}[htb]
    \includegraphics[width=0.5\textwidth,,angle=0,bb=50 85 410 230]{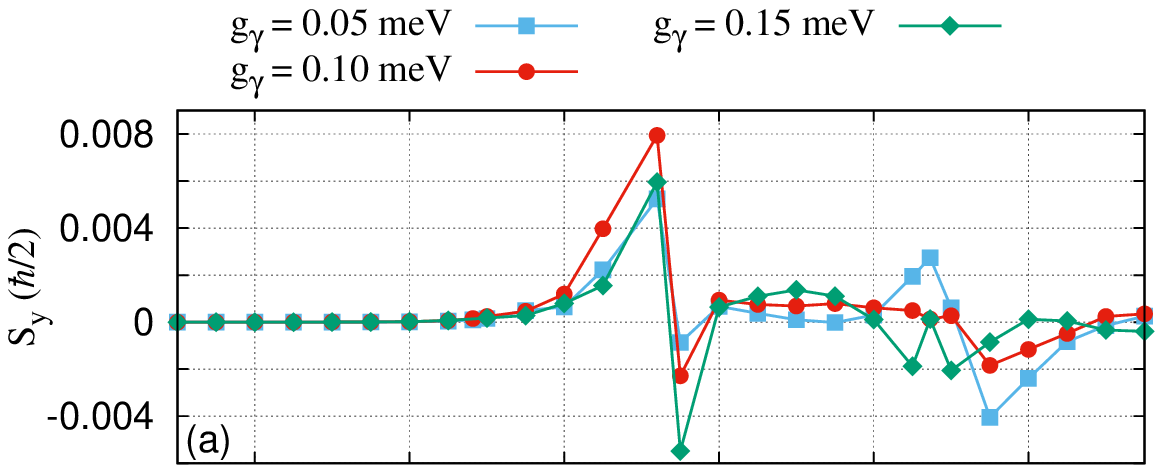}\\
    \includegraphics[width=0.5\textwidth,angle=0,bb=50 85 410 212]{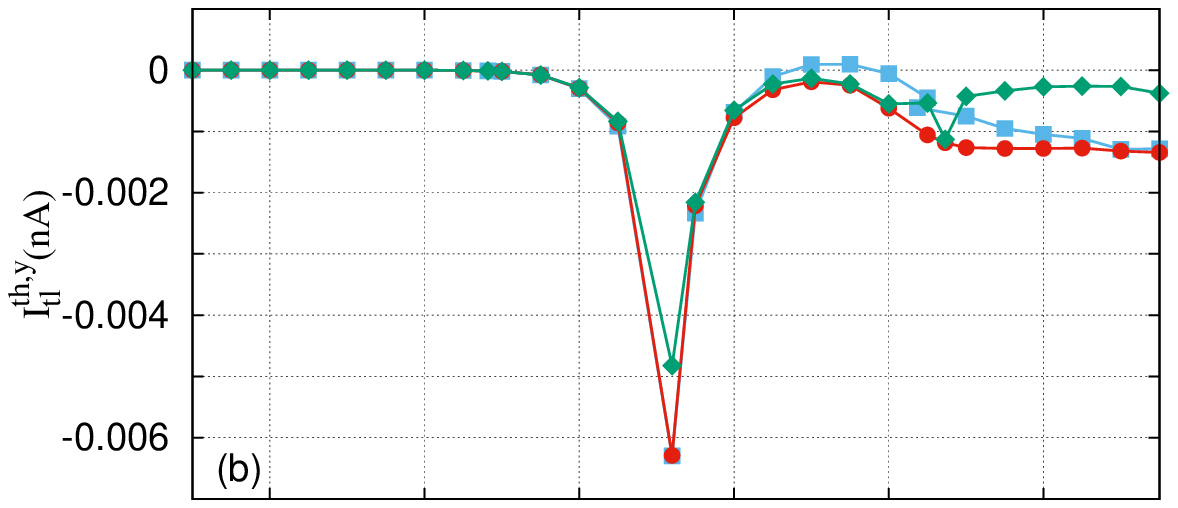}\\
    \includegraphics[width=0.5\textwidth,angle=0,bb=67 60 410 223]{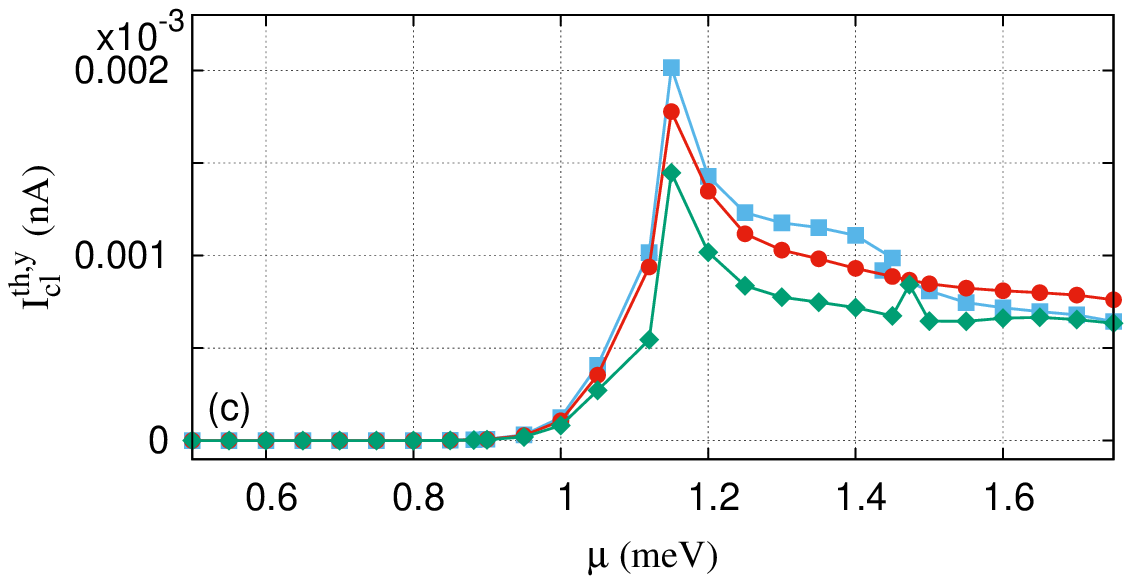}
 \caption{(Color online) (a) $S_y$ spin polarization, (b) TL thermospin polarization current,
 and (c) CL thermospin polarization current versus the chemical potential of 
 the leads $\mu$ for three different values of the 
 electron-photon coupling  strength $g_{\gamma} = 0.05$~meV (blue rectangles),  $g_{\gamma} = 0.1$~meV (red circles), 
 and $g_{\gamma} = 0.15$~meV (green diamonds) with photon energy $\hbar \omega_{\gamma} = 0.55$~meV and 
 linearly polarized photon field in the $x$-direction.  The Rashba coupling constant is $\alpha = 12$~meV nm,
 $B = 10^{-5}$~T, and $\hbar \Omega_0 = 1.0~{\rm meV}$.
 The temperature of the left (right) lead is fixed at $T_{\rm L} = 0.41$~K ($T_{\rm R} = 0.01$~K) implying 
 thermal energy $k_B T_{\rm L} = 0.35$~meV ($k_B T_{\rm R} = 0.00086$~meV), respectively.}
\label{fig07}
\end{figure}

Figure \ref{fig07} demonstrates (a) the $S_y$ spin polarization, (b) TL thermospin polarization current, 
and (c) CL thermospin polarization current as a function of the chemical potential of the leads 
for three values of the electron-photon coupling strength $g_{\gamma}$ in the 
$x$-polarized photon field assuming that the cavity initially contains no photon.

Comparing the system without the photon field shown in \fig{fig05}, the $S_y$ spin polarization is increased 
in the presence of the photon field. The $x$-polarized photon field increases the 
kinetic momentum of the electrons of the QR system in the $x$-direction.
Therefore, tuning the electron-photon coupling strength the 
$S_y$ spin polarization is further enhanced at $\mu = 1.12$~meV (degenerate energy levels).
As the result, the TL thermospin polarization currents are decreased at 
higher value of the electron-photon coupling strength. A reduction in the CL thermospin current is 
observed in the presence of the photon field as well.

\bigskip

\section{Conclusion and Remarks}\label{Sec:IV}

We have studied thermospin properties of a multilevel interacting QR system 
under the influence of a linearly polarized photon field in transport direction. Using a general master equation, 
the spin polarization and thermospin current are calculated.
In the absence of the photon field, but tuning the Rashba coupling constant,
degenerate energy states can be formed leading to 
spin accumulation in the QR system. Consequently, spin polarization can be enhanced and it
minimizes the thermospin current. Under the photon field, photon replica states are formed and 
the photon field increases the kinetic momentum of the electrons. 
As a result, the photon field maximizes the spin polarization and the thermospin current 
is decreased. 
Our results contribute to the expansion of the knowledge accumulating in the 
emerging field of spin caloritronics 
for nanoscale systems.

\bigskip

\begin{acknowledgments}
This work was financially supported by the Research
Fund of the University of Iceland, the Icelandic Research
Fund, grant no. 163082-051, and the Icelandic Infrastructure Fund. 
We acknowledge the University of Sulaimani, Iraq, 
and the Ministry of Science and Technology, Taiwan 
through Contract No. MOST 106-2112-M-239 -001 -MY3
\end{acknowledgments}

%

%

%
%
\end{document}